\documentclass[12pt,a4paper,final]{iopart}
\usepackage{iopams}  
\usepackage{graphicx}
\usepackage[breaklinks=true,colorlinks=true,linkcolor=blue,urlcolor=blue,citecolor=blue]{hyperref}

\begin{document}

\title[Micro and macrophase separation in discrete potential fluids]{Micro and macrophase separation in discrete potential fluids}

\author{I. Guill\'en-Escamilla*, J. G. M\'endez-Berm\'udez,  J.C. Mixteco-S\'anchez}
\address{Departamento de Ciencias Naturales y Exactas, CUValles, Universidad de Guadalajara, Km 47.5 Carretera Guadalajara-Ameca, 46600, Ameca, Jalisco, Mexico.}
\ead{ivan.guillen@academicos.udg.mx*}

\author{and G. Arlette. M\'endez-Maldonado}
\address{Departamento de F\'{\i}sica, CUCEI, Universidad de Guadalajara, Blvd. Marcelino Garc\'{\i}a Barrag\'an 1421, esq Calzada Ol\'{\i}mpica, 44430, Guadalajara, Jalisco, Mexico.}

%\ead{author.two@mail.com}

\begin{abstract}
We study the liquid-vapor phase diagram and the structural properties of discrete potential fluids by means of Gibbs ensemble simulations and integral equations theory. We consider three discrete fluids, namely, the square well (SW), the square well-barrier (SWB), and the square well-barrier-well (SWBW). They represent simple models for fluids with competing interactions that may exhibit a rich micro and macroscopic phase behavior depending on both strength and range of the attractions and repulsions in the potential. Here we emphasize the structural behavior near the liquid-vapor coexistence. Our findings indicate that for the SWB fluid a possible scenario of a microscopic phase separation, associated to a cluster-like formation near the critical region, is observed and could be understood as a frustration mechanism of the liquid-vapor transition when either the strength or the range of the repulsion increases. This microscopic-like separation can be inhibited by suppressing the repulsion or adding an additional well to the interaction potential. However, in the SW fluid of long-range potential, we have found results that point out towards evidence of a microscopic aggregation driven purely by attractions.

\end{abstract}

\vspace{2pc}

%\submitto{\JPCM}
%\maketitle

\section{Introduction}

During the last few years, it has been demonstrated that a large variety of homogeneous and inhomogeneous phases in equilibrium and out of it, in both simple and complex fluids, can be tuned only by changing the range and strength of the interaction potential \cite {expertl1, expert9, expertl5, expert8,Lu2008}. This results from the superposition of the repulsive and attractive contributions in the interaction potential, i.e., the physical behavior of the observed phases emerges from the competition among different types of interactions that lead to complex potentials between particles of the fluid \cite{AStradner, AShukla_c1, AStradner2, Bomont, Imperio2004,Imperio2006,Verso, Archer1, Archer2, Archer2008,Kahl2010,Archer3, Archer4, Ghezzi,Sear,Chakrabarti,Min,Costa2011,SchollPaschinger}. Competing interactions have been extensively used to investigate, for example, the formation of ordered structures in complex systems, such as globular protein solutions \cite{AStradner,AShukla_c1, AStradner2}, the effective interactions between solute particles in a subcritical solvent \cite{Chakrabarti}, the temperature dependence of the cluster-like formation in double-Yukawa fluids \cite{Bomont,Min,Costa2011,Bomont2012} and the so-called microphase separation in two- and three-dimensional systems \cite{Imperio2004,Imperio2006,Verso, Archer1, Archer2, Archer2008,Kahl2010,Archer3, Archer4}.

From the structural point of view, a macroscopic or thermodynamic phase separation in monodisperse fluids can be represented by means of the divergency of the static structure factor, $S(q)$, at $q = 0$, i.e., in the long wavelength limit \cite{Bhatia}. In contrast, a microphase separation refers, typically, to the presence of a peak in the $S(q)$ at wavelengths $q\le q_{m}(\equiv2\pi/d)$, where $d$ is either the particle diameter or the meaninterparticle distance. For further details of the definition of a microphase separation, see, e.g., work done by A. Archer and co-workers \cite{Archer1, Archer2, Archer2008,Kahl2010,Archer3, Archer4}. This characteristic peak is clearly associated with a kind of particle aggregation \cite{AStradner}, however, there still exists a debate on whether such a peak is an indicator of a correlation between aggregates, i.e., clusters, in the fluid or it represents an intermediate range order structure, see, e.g., \cite{Liu2011} and references therein. Nonetheless, the degree of ordering of the aggregate should be a function of the peak height, but, unfortunately, this parameter does not provide explicit information about the kind of ordering and the possible transition from intermediate to permanent order. Further studies in this direction will be discussed elsewhere \cite{Doug2012}.

Hence, in this work, the term \textit{microphase separation} is simply used to highlight the presence of an additional peak in the $S(q)$ at low-$q$ values, which is driven by the competition between the attractive and repulsive contributions of the interaction potential. Commonly, a continuous interaction potential that takes into account a short-range attraction and a long-range repulsion, i.e., a double-Yukawa potential, is used to represent a large variety of systems with competing interactions \cite{Bomont, Imperio2004,Imperio2006,Verso, Archer1, Archer2, Archer2008,Kahl2010,Archer3, Archer4,Min,Costa2011}. Here we follow a different strategy and consider a potential represented by a superposition of square-wells and square-barriers. A fluid where particles interact with this kind of potential is usually known as a discrete potential fluid (DPF) \cite{expert03}. The importance of DPFs resides in the fact that they allow us to study separately the effects produced by the different attractive and repulsive components of the potential \cite{expertDCP, expert04, expert12a}; an aspect that cannot be done with continuous potentials, where only global effects can be distinguished. Moreover, thermodynamic and structural properties of DPFs have been studied by employing computer simulations, perturbation-like theories, and integral equations theory \cite{expert7,expertl4,expertDelrio1,expert23,expert34,expert15,expert21,expertTW,expert41,expert28,expert12a,Jin2011}. Furthermore, it is shown that DPFs exhibit multiple phase transitions \cite{expertDCP, expert04}, and analytical expressions for direct correlation functions have been developed \cite{Hlushak2009,Guillen2010,Guillen2011,Hlushak2011}, which, for example, could be used as new reference systems in perturbation-like theories or incorporated in dynamical approaches, see, e.g., \cite{Ramirez2010, OlaisGovea}, to account for the diffusive process in fluids with competing interactions. Hence, DPFs are ideal candidates to have a full control on the strength and range of all the contributions in the interaction potential.

Recently, we have studied the structure far from the coexistence region of three types of DPFs, namely, the square well (SW), the square well-barrier (SWB), and the square well-barrier-well (SWBW) \cite{expert12a}. We have found the following interesting features. The inclusion of attractive and repulsive components in the potential promotes changes in the local structure and long-range order in the fluid. In particular, the attractive components induce higher compressibility. In addition, we elucidated the possible formation of clusters or domains, but this point was not fully reviewed due to the appearance of aggregates that does not occur frequently at high temperatures, however, we observed that some clusters locally exhibit a fluid-like order, whereas the repulsive part tends to stabilize the fluid, inhibiting the formation of these domains and lowering the compressibility  \cite{expert12a}. Those properties suggest a richer structural behavior when the fluid approaches to the critical region. Therefore, both phase behavior and the structure near the coexistence region are discussed in the present work. In particular,  we study the influence of the interaction potential parameters, i.e., strength and range, on the macro and microphase separation in DPFs.

The study of the liquid-vapor phase equilibrium is done using the so-called Gibbs ensemble Monte Carlo (GEMC) method \cite{expertFrenkel, expertDelRio2002} and the critical temperature and density are obtained through the law of rectilinear diameters \cite{expert10} and the scaling law described in \cite{expert12}. The structural properties are investigated by means of the numerical solution of the Ornstein-Zernike (OZ) equation \cite{OZ}. We have solved such an equation by using different closure relations, namely, the Percus-Yevic (PY) \cite{PY}, meanspherical approximation (MSA) \cite{MSA}, hypernetted-chain (HNC) \cite{HNC}, hybrid meanspherical approximation (HMSA) \cite{HMSA}, and Rogers-Young (RY) \cite{RY}, in order to monitor their limits of applicability to the study of the microstructure of DPFs and to establish what closures are suited in predicting reasonably well the microphase separation in fluids with competing interactions. Thus, we compare our theoretical predictions with MC computer simulations in the canonical ensemble.

After the Introduction, the paper is organized as follows. In section II we briefly describe the discrete interaction potential used to model the DPF, the simulation technique and the integral equations theory. Section III deals with the liquid-vapor phase coexistence of the aforementioned DPFs and in section IV we show the main results of their structural behavior. This work ends with a section of concluding remarks.

\section{Discrete interaction potential, integral equations theory and computer simulations} 

\begin{figure}[htbp]
  \centering
  \includegraphics[width=10cm,height=7.5cm]{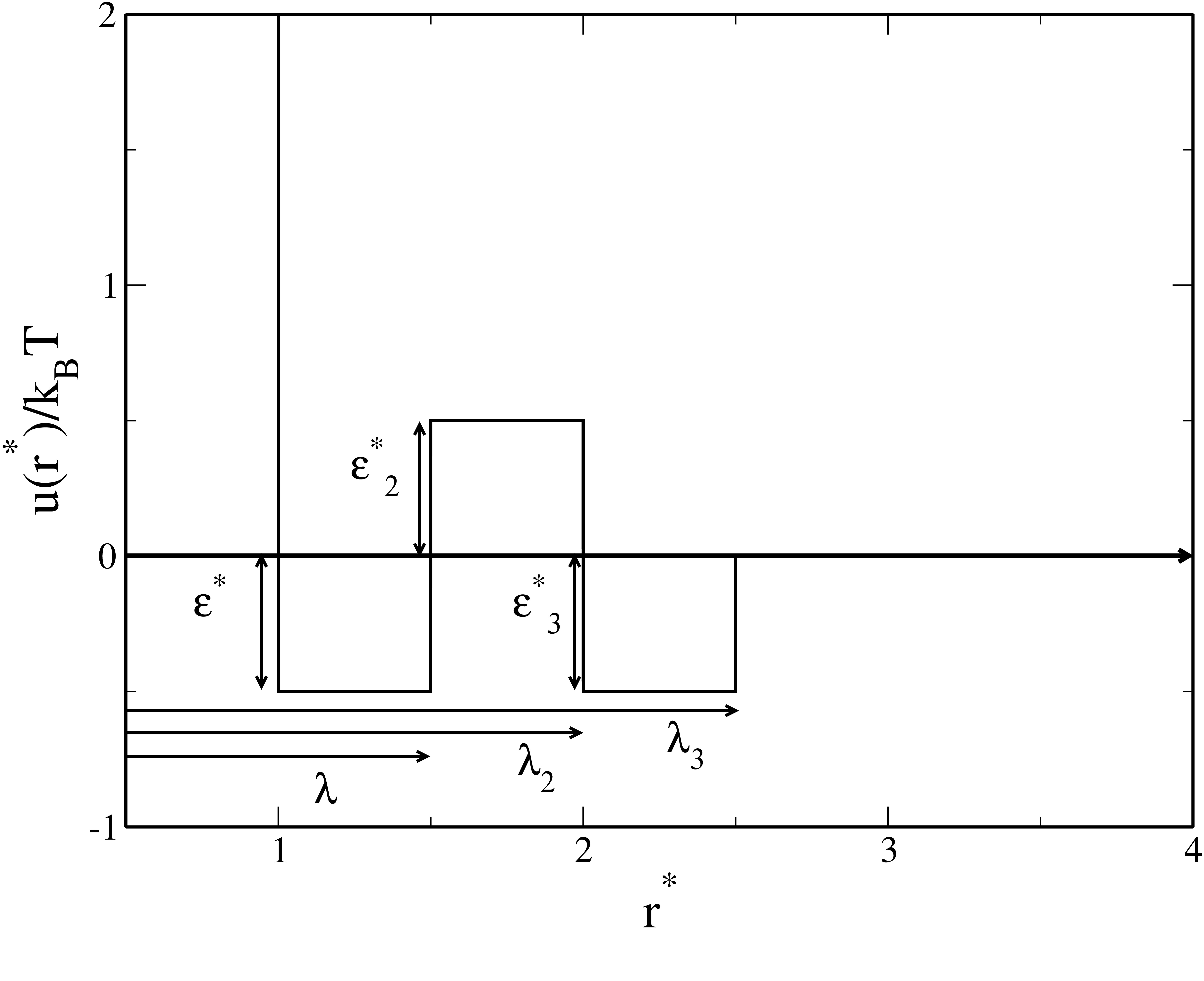}
\caption{Representation of the discrete potential used in this work, where $r^{*}\equiv r/\sigma$.}
\label{fig1}
\end{figure}

Our system is made up of spherical particles of diameter $\sigma$ and it is in thermal equilibrium at absolute temperature $T$. Particles interact through a discrete potential that sequentially includes a hard-sphere, a square-well, a square barrier, and a second square-well \cite{expert12a}. This pair potential has the following analytic representation \cite{expert12a}
\begin{eqnarray}
\label{eq2}
u(r) = \left\{ 
\begin{array}{cc}
\infty, & \mbox{$r<\sigma$}; \\ 
-\epsilon, & \mbox{$\sigma \le r < \lambda\sigma$}; \\ 
\epsilon_{2}, & \mbox{$\lambda\sigma \le r < \lambda_{2}\sigma$}; \\
-\epsilon_{3}, & \mbox{$\lambda_{2}\sigma \le r < \lambda_{3}\sigma$}; \\
0, & \mbox{$\ge\lambda_{3}\sigma$}, 
\end{array}
\right.
\end{eqnarray}
where the parameters $\lambda$, $\lambda_{2}$, and $\lambda_{3}$ define both the discontinuity points and the range of the attractive and repulsive contributions in the potential. On the other hand, parameters $\epsilon$, $\epsilon_{2}$, and $\epsilon_{3}$, characterize the strength of such contributions. A schematic representation of the potential is shown in figure (\ref{fig1}). 

The depth of the first well, $\epsilon$, and the particle diameter are used to express the reduced temperature $T^{*} = k_{B} T /\epsilon$ and the reduced density $\rho^{*}=\rho \sigma^{3}$, respectively, being $k_{B}$ the Boltzmann's constant and $\rho$ the particle number density. From equation (\ref{eq2}), several cases can be considered depending on the values chosen for the interaction parameters. For example, equation (\ref{eq2}) reduces to the well-known hard-sphere (HS) potential when $\epsilon=\epsilon_{2}=\epsilon_{3} = 0$. We will consider the three following cases of equation (\ref{eq2}): (a) the SW potential, defined by $\epsilon > 0$ and $\epsilon_{2}=\epsilon_{3} = 0$; (b) the SWB potential, defined by $\epsilon > 0$, $\epsilon_{2} > 0$, and $\epsilon_{3} = 0$; and (c) the SWBW potential, defined by the full equation (\ref{eq2}). In this way, SW fluids are characterized by a single well, SWB fluids by a well and a barrier, and SWBW fluids by two wells with a barrier in between.

The SW fluid is, probably, the most studied DPF \cite{Guillen2010, expertDelrio1,expert28,expert12a,Hlushak2009, expert03, expert16,   expert31, expert33,expert34, expert35, LopezHaro, SolanaAkhouri,  expert36, expert37,expert41, Jin2011, expert23,expert15, expert21,expertl4,expert7}. Here we review its thermodynamic and structural properties, since it will be considered as a reference fluid in our study. In particular, we explore its physical properties in an interval $\lambda \in [1.5,3]$ in steps of $\Delta \lambda =0.5$. We point out that the case $\lambda=1.5$ is a typical value chosen within the context of simple liquids, while the case  $\lambda>2.0$ has been less studied.

We have extensively studied different values for $\lambda_{2}$ and $\epsilon_{2}$ of the SWB fluid, however, we show only those cases where a clear competition between micro and macrophase separations seems to be observed. More precisely, we consider the following parameters for the SWB fluid: $\lambda=$1.25 and 1.5;  $\lambda_{2}=$  2.0 and 3.0; $\epsilon_{2}\in [0.1,0.5]$. In the case of the SWBW fluid the potential parameters used are: $\lambda=1.5$, $\lambda_{2}=2$, $\lambda_{3}=2.5$, $\epsilon_{2} \in [0,0.5]$ and $\epsilon_{3} \in [0,0.5]$.

In our Gibbs ensemble simulations \cite{expertFrenkel,expertl2}, we initially place 2048 particles uniformly distributed into two cubic boxes of equal volume. Then, we carry out the following trial moves: displacement of particles in each box, change of volume while the total volume remains constant, and particle exchange between the boxes; a Monte Carlo cycle consists of randomly performing the previous operations at the ratio 800:199:1, respectively. We use $2.5\times 10^5$ Monte Carlo cycles to equilibrate the system and $2.5\times 10^5$ production cycles. The acceptance ratio for both particle displacement and volume changes is fixed to $50\%$. 

The phase diagram of the SW fluid is compared with results obtained using the so-called self-consistent Ornstein-Zernike approximation (SCOZA) and taken from Ref. \cite{expert28}. In those SWB fluids where a liquid-vapor phase transition appears, we study the microstructure near, but above, the critical point by means of MC computer simulations in the $NVT$ ensemble. We use $2048$ particles, $2\times 10^5$ cycles to equilibrate the system, $2\times 10^5$ to obtain statistics and $50\%$ of acceptance. Moreover, when the interaction is sufficiently long-range the phase coexistence disappears. In this case, we focus on the microstructure at low temperatures.

From  the theoretical point of view, the microstructure can be studied by means of the solution of the well-known Ornstein-Zernike (OZ) equation \cite{OZ}, which defines the direct correlation function, $c(\vec{r})$, in terms of the total correlation function, $h(\vec{r})=g(\vec{r})-1$, where $g(\vec{r})$ is the radial distribution function. For homogeneous and isotropic fluids, the OZ equation takes the form \cite{OZ},
\begin{equation}
\label{OZ}
h(r)=c(r)+\rho \int c(\vec{r}')h(|\vec{r}-\vec{r}'|)d\vec{r}'.
\end{equation} 
To solve equation (\ref{OZ}), one needs a relation between $c(r)$ and $h(r)$ that, on one hand, incorporates the information of the interaction potential and, on the other hand, allows us to close the set of equations. The most general closure relation can be written as \cite{Hansen}
\begin{equation}
\label{cr}
c(r)=e^{-\beta u(r)+ \gamma(r)+ B(r)}-\gamma(r)-1,
\end{equation} 
where $\gamma(r)=h(r)-c(r)$ is called the indirect correlation function and $B(r)$ is the bridge function \cite{Hansen}. However, generally speaking, the form of $B(r)$ is unknown and further approximations of it are needed to find the solution of the OZ equation. In this work, we numerically solve equation (\ref{OZ}) by employing the Ng method \cite{Ng} and assuming a particular choice for $B(r)$. We use different closure relations to determine, in terms of the potential parameters, the regime of applicability where they describe accurately the microstructure of the fluid. In addition, one of us has recently discussed on the importance of thermodynamic self-consistency to describe cluster-like correlations in double-Yukawa fluids \cite{Min}. Then, we use both types of closure relations, namely, those which do not guarantee the thermodynamic self-consistency, i.e., PY, MSA and HNC  \cite{PY,MSA,HNC}; and the ones which incorporate it partially, i.e., HMSA and RY \cite{HMSA,RY}. They can be easily expressed in terms of $B(r)$ as follows,
\\

PY
\begin{equation}
\label{PY}
B(r)=\ln{\left[1+\gamma(r)\right]}-\gamma(r),
\end{equation}

MSA
\begin{equation}
\label{MSA}
B(r)=\ln{\left[1+\gamma(r)-\beta u(r)\right]}+\beta u(r)-\gamma(r),
\end{equation} 

HNC
\begin{equation}
\label{HNC}
B(r)=0,
\end{equation} 

RY
\begin{equation}
\label{RY}
B(r)=\ln{\left[{1+\frac{e^{(\gamma(r))f(r)}-1}{f(r)}}\right]}-\gamma(r),
\end{equation}

HMSA
\begin{equation}
\label{HMSA}
B(r)=\ln{\left[{1+\frac{e^{(\gamma(r)-\beta u_{a}(r))f(r)}-1}{f(r)}}\right]}+\beta u_{a}(r)-\gamma(r),
\end{equation} 
where $u_{a}(r)$ is the attractive contribution to the interaction potential $u(r)=u_{r}(r)+u_{a}(r)$, with $u_{r}(r)$ as the repulsive contribution. In equations (\ref{RY}) and (\ref{HMSA}), $f(r)$ is a mixing function defined as $f(r)=1-\exp(-\alpha(r))$, $\alpha$ being the mixing parameter. The latter is calculated by demanding thermodynamic self-consistency, which is reached by equating the isothermal normalized compressibility $\chi$ of the fluid from the virial route, ${{\chi}_{v}^{-1}}=\left(\frac{\partial{{\beta}P}}{\partial\rho}\right)_{T}$, and the compressibility route, ${{\chi}_{c}^{-1}}={1-\rho\tilde{c}(q=0)}$, i.e., ${{\chi}_{c}^{-1}}={{\chi}_{v}^{-1}}$ \cite{RY,HMSA}, where $P$ is the pressure of the system and $\tilde{c}(q)$ is the Fourier transform of the direct correlation function. Thus, by choosing one of the above closure relations, we insert it into equation (\ref{cr}) in order to solve the OZ equation (\ref{OZ}).

We do not report the excess chemical potential, $\mu$, but it can be easily determined during the simulations without additional cost by simply evaluating the expression \cite{expertFrenkel}
\begin{equation}
\mu_{1}= k_{B}T \ln\frac{1}{\Lambda}\left<\frac{V_{1}}{n_{1}+1}\exp(\beta \Delta u) \right>,
\end{equation}
where $\Delta u$ is the energetic cost of inserting a particle in box 1, $\Lambda$ is the de Broglie thermal length, and $<\cdots>$ stands for an ensemble average. Similarly, $\mu_{2}$ in the second box can be straightforwardly evaluated. We also can extract the pair correlation function and the pressure at coexistence. Both quantities can then be used to test theoretical predictions, as those based on the liquids theory, such as the Ornstein-Zernike equation (\ref{OZ}) \cite{OZ}. 

The equation of state of discrete potential fluids described by equation (\ref{eq2}) can be straightforwardly determined from the virial equation \cite{expert12a},
\begin{equation}
\frac{\beta P}{\rho}=1+\frac{2}{3}\pi \rho \sigma^{3}\left[\sum^{3}_{i=0}\lambda_{i}^{3}\Delta g(\lambda_{i}) \right],
\end{equation}                                                              
where $\lambda_{i}$ denotes the discontinuity points of the potential, and $\Delta g(\lambda_{i})=g(\lambda^{+}_{i})-g(\lambda^{-}_{i})$ is the difference in contact values of the radial distribution function at the discontinuity. For a HS fluid $i=0$ and the summation only contains the term $\Delta g(\sigma) = g(\sigma^{+})$. The SW fluid additionally involves 
$\Delta g(\lambda) = g(\lambda^{+})-g(\lambda^{-})$, the SWB also considers the term $\Delta g(\lambda_{2})=
g(\lambda^{+}_{2})-g(\lambda^{-}_{2})$, and, finally, the SWBW includes the contribution $\Delta g(\lambda_{3})=
g(\lambda^{+}_{3})-g(\lambda^{-}_{3})$.

\section{Liquid-vapor phase diagram}

\subsection{SW fluid}

\begin{figure}[htbp]
  \centering
  \includegraphics[width=10cm,height=8cm]{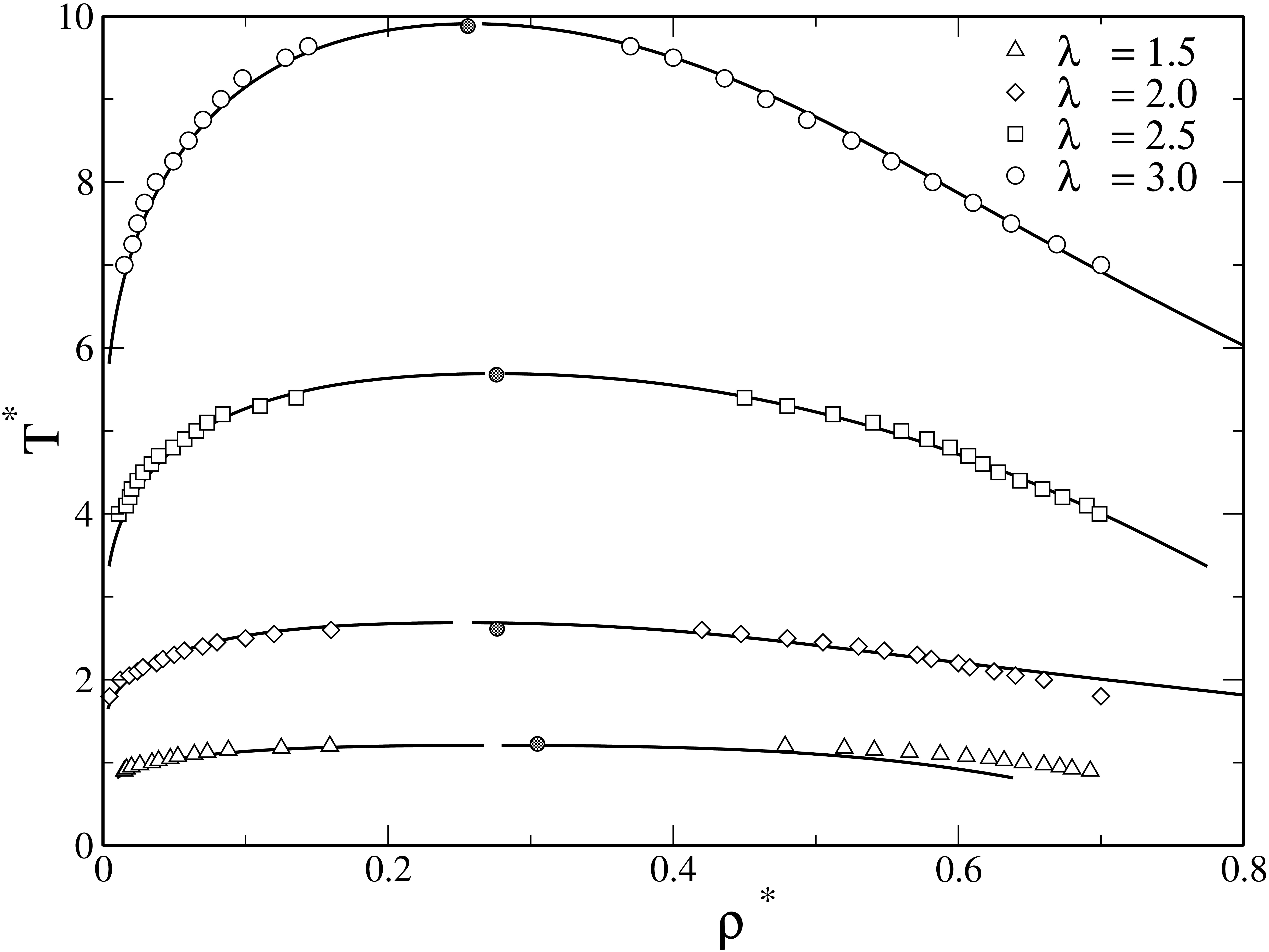}
\caption{Liquid-vapor phase diagrams of SW fluids for $\lambda=1.5, 2, 2.5$ and $3$. 
Open symbols correspond to simulation data and lines to SCOZA results taken from Ref. \cite{expert28}. The closed circles indicate the simulation critical points.}
\label{fig2}
\end{figure}

Although the phase coexistence of the SW fluid has been studied in detail \cite{expert00, expert03, expert16, expert19, expert20,expert22,expert3,expert17,expert5,expert31, expert32, expert33,expert34, expert35, expert36,expert37,expert38, expert39,expert40,expert41}, for the sake of the discussion, we revisit its dependence on $\lambda$, which is displayed in figure (\ref{fig2}). We show different attraction ranges; $\lambda = 1.5, 2.0, 2.5, 3.0$, i.e., from short to long range attractions. We observe that an increase in $\lambda$ gives rise to higher coexistence temperatures. From a simple physical perspective, this fact can be explained as follows. When the thermal energy is not large enough as compared with the well depth, particles, on average, will be located in the region of the attractive well to minimize the system free energy. This can be the driving force that permits the formation of well-defined particle domains, as already discussed in \cite{expert12a}. Then, the domain formation is favored with an increase in $\lambda$, as we will see further below. Thus, to avoid such agglomeration of particles and find the system within the fluid phase, the thermal energy has to increase as well. Moreover, we observe that the critical density does not change significantly with the interaction range. Additionally, our simulation data agree very well with previous estimations reported in the literature using molecular dynamics and Gibbs ensemble Monte Carlo \cite{expert4,expert6} (data not shown). We also find a good agreement between simulations and those results recently predicted with SCOZA (solid lines) \cite{expert28}.

%%%%%%%%%%%%%%%%%%%%%
%   Figure 2
%%%%%%%%%%%%%%%%%%%%%

\subsection{SWB fluid}

We now discuss our results for the SWB fluid. To better understand its phase coexistence behavior, we conveniently define the following quantities: $\epsilon^{*}_{r}\equiv\epsilon_{2}/\epsilon$ and $\lambda^{*}_{r}\equiv (\lambda_{2}-\lambda_{1})/\lambda_{1}$, which describes the strength and range of the repulsion, respectively, in units of the reduced range, $\lambda$, and depth, $\epsilon$, of the well. $\epsilon$ is our energy unit, its reduced value is always $\epsilon^{*}=1$. 

\begin{figure}[htbp]
  \centering
  \includegraphics[width=10cm,height=8cm]{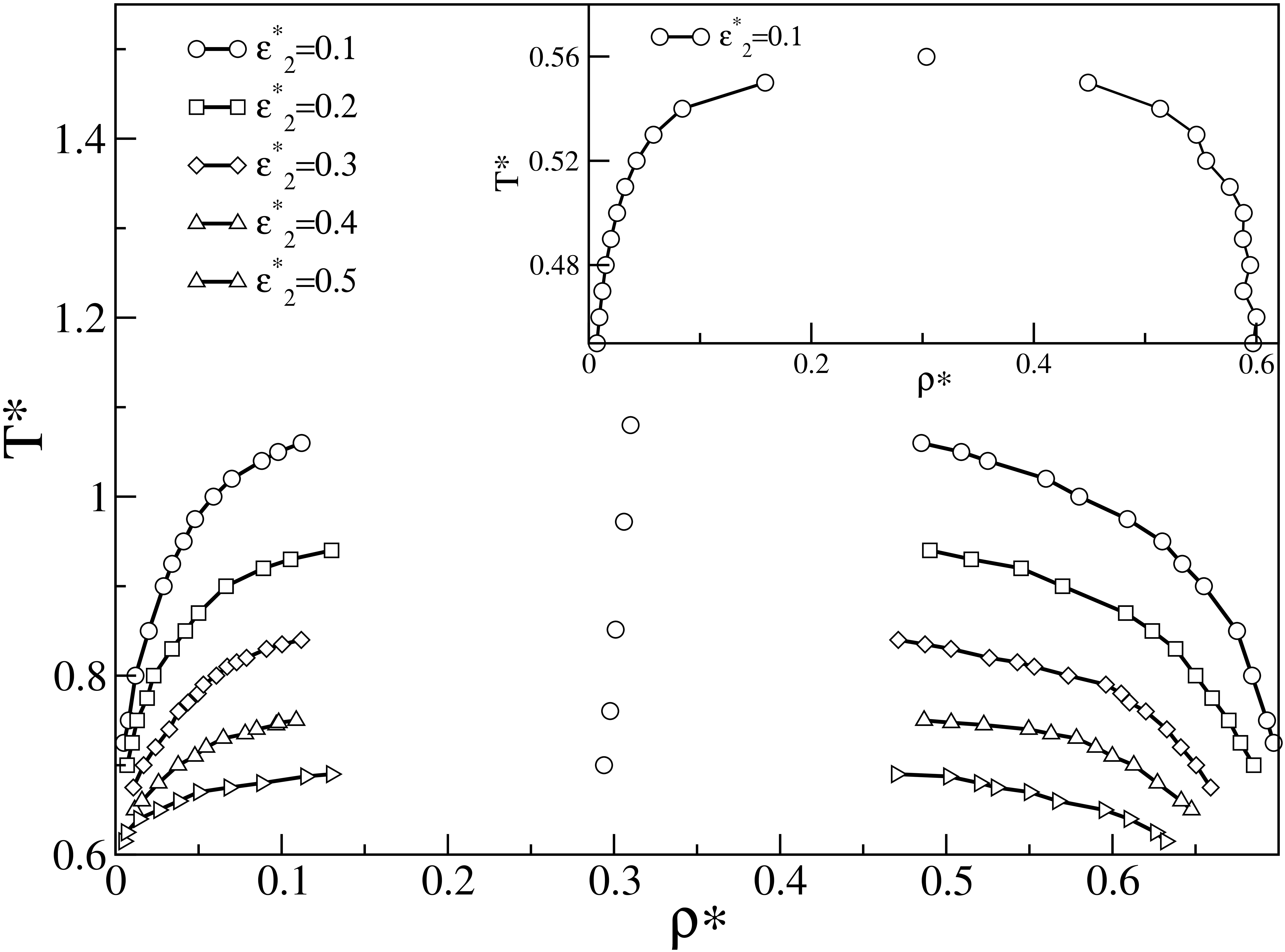}
\caption{Liquid-vapor phase diagrams obtained with Gibbs ensemble simulations of SWB fluids for $\lambda=$1.5,  $\lambda_{2}=2.0$, $\epsilon^{*}=$1.0, 
varying $\epsilon^{*}_{2}=$ 0.1, 0.2, 0.3, 0.4 y 0.5. In the inset, the liquid-vapor phase diagram for $\lambda=1.25$ and $\lambda_{2}=2$ is shown, only appears for $\epsilon \le 0.1$.The circle indicates the critical point and lines are just a guide for the eye.}
\label{fig5}
\end{figure}

In figure (\ref{fig5}) we study the effect of the repulsion strength, $\epsilon^{*}_{2}=\epsilon^{*}_{r}$, on the phase behavior. We use the following parameters for the potential: $\lambda= 1.5$ and $\lambda_{2} = 2$, i.e., $\lambda^{*}_{r}\approx 0.33$. We observe interesting effects induced by the repulsion, the critical values of the density and temperature lower as the $\epsilon_{2}$ values increase. The barrier tends to inhibit the phase transition; our exploration through simulations indicates that the phase transition is completely inhibited for $\epsilon^{*}_{2} > 0.8$. 

This can be understood as follows: we can notice that when increasing the height barrier, the energy difference $\Delta E_{t}  = \epsilon_{2}  + \epsilon$ increases as well, thus forming an effective barrier that avoid that particles lie out the potential well, but also the increment of $\epsilon_{2}$ impedes that the particles lie in the potential well, due to the thermal fluctuations and, therefore, it is necessary a thermal energy large enough so that the particles can be located out the well, this promotes the critical point to lower as the height of the barrier increase.

%This can be understood as follows: we can notice that when increasing the height barrier, the energy difference $\Delta E=\epsilon_{2}+\epsilon$ as well increases, thus forming an effective barrier that tries to impede that particles lie in the potential well due to the thermal fluctuations and, therefore, to promote the phase separation. One might be interested in finding the threshold effective barrier, $\Delta E_{t}$, that inhibits completely the liquid-vapor transition. A simple approximation should consider that $\Delta E_{t}$ has to be slightly greater than the thermal energy at which the phase coexistence of the pure SW fluid appears of range $\lambda$, i.e., at the critical temperature, $T^{*}_{SW}(\lambda)$. Thus, we can assume, for simplicity, that such threshold barrier satisfies the relation $\Delta E^{*}_{t}\approx\frac{3}{2}T^{*}_{SW}(\lambda)$. Then, according to figure (\ref{fig2}), one has that $T^{*}_{SW}(\lambda=1.5) \approx 1.2$. Therefore, the phase coexistence in the SWB fluid is suppressed at $\epsilon_{2}^{*} =0.8$; a value that agrees completely with our simulation results.

%
%%%%%%%%%%%%%%
%   Figure 5
%%%%%%%%%%%%%%
%

One might be interested in finding the threshold effective barrier, $\Delta E_{t}$, that inhibits completely the liquid-vapor transition. Within our exploration, we find that it does not depend only on the height of $\epsilon_{2}$ but also on the repulsive range. For a short width repulsive interaction,  $\Delta \lambda= \lambda_{2}-\lambda\lesssim \sigma/2$,  a higher barrier is needed to inhibit the coexistence, for example, if $\lambda=1.5$ and $\lambda_{2}=2.0$ a barrier height of $\epsilon_{2}=0.8$ is capable of inhibiting the LV coexistence. However, for $\Delta \lambda= \lambda_{2}-\lambda \gtrsim \sigma/2$, a small barrier is needed in order to promote the cluster phase formation and inhibit the coexistence. In the inset of the figure 3, is shown the diagram phase for the attractive and repulsive interaction short-range and long-range, respectively; when $\lambda=1.25$ and $\lambda_{2}=2$ we find that the coexistence appear if $\epsilon_{2}\le 0.1$. In the next section, the study of this behavior is extended through the structure properties.

Additionally, we observed that when the height of the barrier decreases below zero, the phase diagram shifts to higher temperatures (data not shown), since the barrier becomes a second well and particles can be distributed in both wells strongly favoring the phase transition. Besides, from theoretical point of view, it is quite difficult to evaluate the liquid-vapor phase diagram of the SWB using advanced approaches, such as SCOZA, since in that particular case, the standard procedure described in \cite{expert28} fails to converge numerically when an additional barrier, apart from the hard-core, is explicitly considered in the interaction potential. However, further considerations allow us to evaluate the fluid coexistence of DPFs using SCOZA. This point will be discussed elsewhere.

\subsection{SWBW fluid}

In this case, a secondary attractive well is added to the SWB potential. As before, we maintain the range of wells and barrier fixed at $\lambda =$1.5, $\lambda_{2} =$2.0 and $\lambda_{3} =$ 2.5. For investigating the coexistence phase of the competing potential fluid, we analyze two cases: the increase of the repulsive contribution and the increase of the attractive strength of the second well; for this reason, we present results for the following sequences, one sequence where $\epsilon_1=1.0$ and $\epsilon_{3}=0.1$ are fixed and the height of the barrier is varied, and another sequence where $\epsilon_{1}=1.0$ and $\epsilon_{2}=0.5$ are fixed and $\epsilon_3$ is varied.

\begin{figure}
  \centering
  \includegraphics[width=10cm,height=8cm]{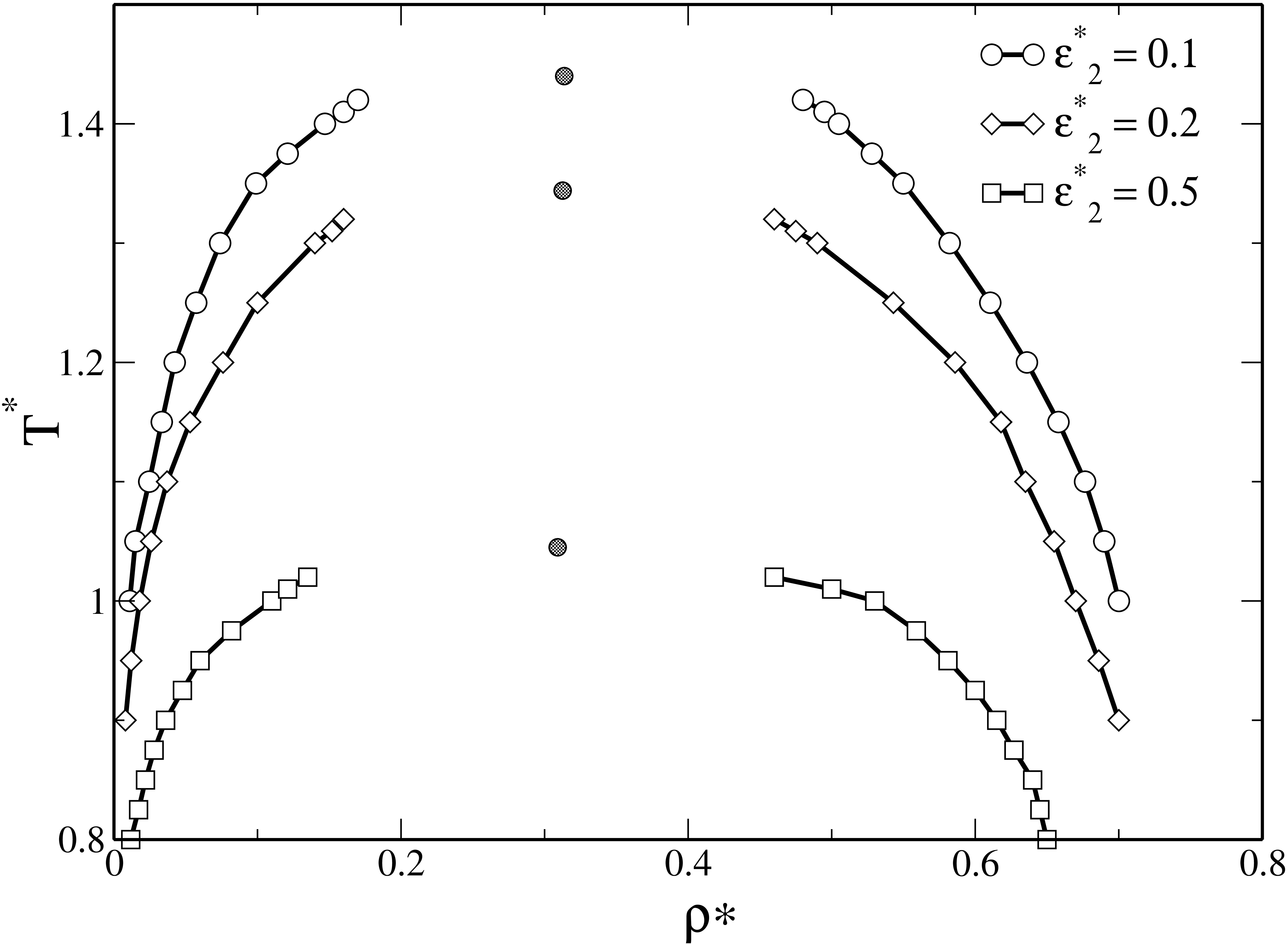}
  %\hspace{0.2cm}
  %\includegraphics[width=6.5 cm]{figure10.eps}
\caption{Phase diagram for SWBW fluids,  the values of  $\epsilon=1.0$ and $\epsilon_{3}=0.1$ are fixed, and the effect of  increase of  the barrier height is studied; the values of the barrier height are: $\epsilon_{2}=$ 0.1, 0.2 and 0.5.  The results were obtained through GEMC and the dark circle indicates the critical point. }
\label{fig7}
\end{figure}

In figure (\ref{fig7}) we show the results of the first case, where the height of the barrier is increased, we observe the same behavior as in the case of the SWB fluid, i.e., the repulsive contribution moves the coexistence region towards lower temperatures when the barrier height increases. However, in comparison with the SBW fluid coexistence the values of temperature at which the coexistence is observed are higher. This is, without a doubt, due to the presence of the second potential well.  The attractive contribution in the potential favors the liquid-vapor coexistence, for this reason the coexistence region appears at higher temperatures than in the SWB case, the second well increases the attractive effective range. On the other hand, the barrier and the second well generate an effective barrier of height $\epsilon_{2}+\epsilon_{3}$ which inhibits the coexistence. As the barrier height is increased the system needs more energy to reach the coexistence, which is reflected in a lower critical temperature as the barrier increases.

%%%%%%%%%%%%%%%%%%%%%%
%   Figure 7
%%%%%%%%%%%%%%%%%%%%% 

\begin{figure}
  \centering
  \includegraphics[width=10.0cm,height=8cm]{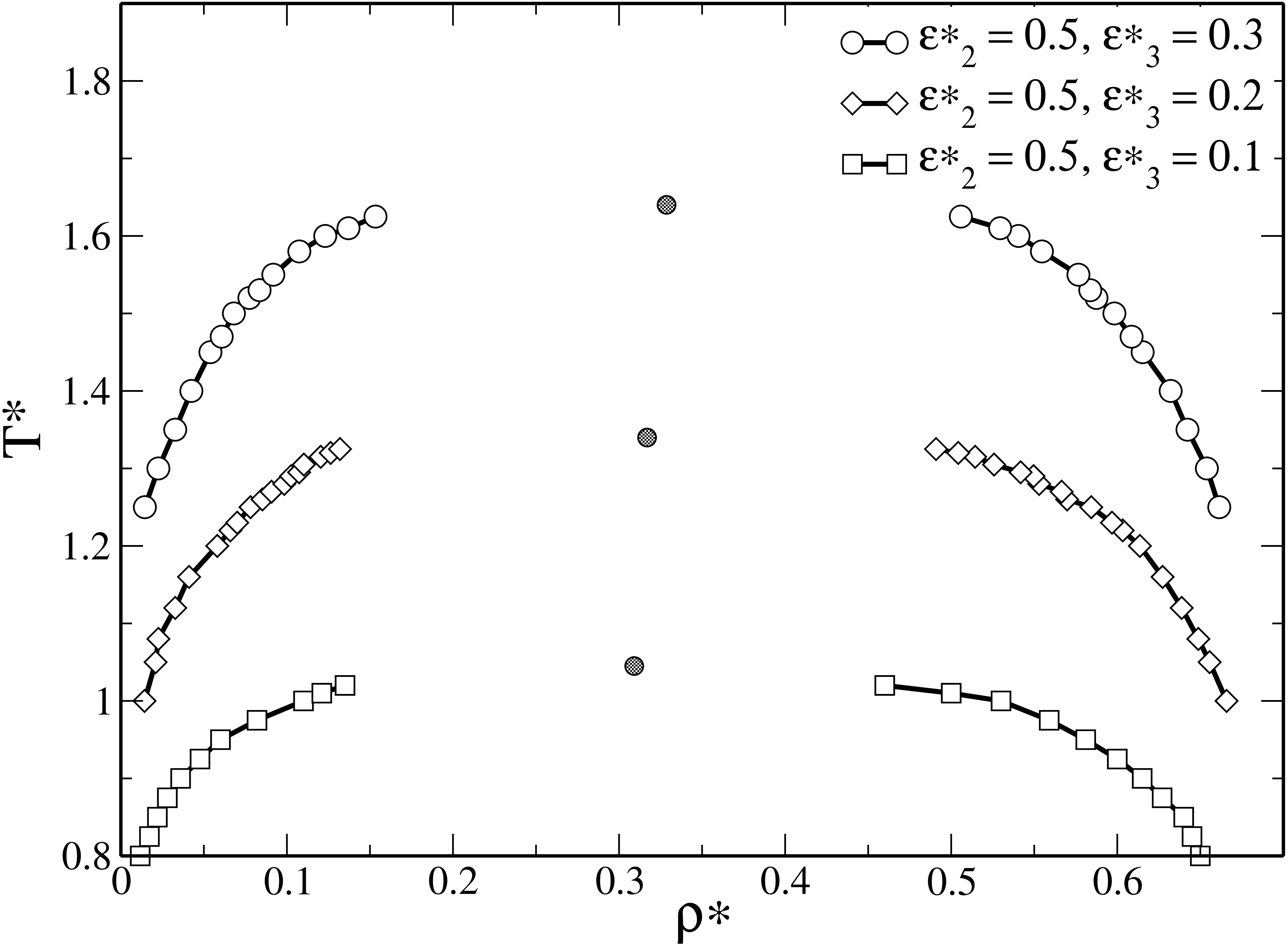}
\caption{Phase diagram for SWBW fluids,  the values of  $\epsilon=1.0$ and $\epsilon_{2}=0.1$ are fixed, and the effect of  increase of  the well depth is studied; the values of the well depth are: $\epsilon_{3}=$ 0.1, 0.2 y 0.3.  The results are computed through GEMC and the dark circle indicates the critical point. }
\label{fig8}
\end{figure}

In order to examine the effect of the second well, we consider three values for the depth $\epsilon_{3}=0.1, 0.2$ and $0.3$. In this case, the region of coexistence is moved towards larger temperatures as the well depth is increased, as we shown in the figure \ref{fig8}, this indicates that the second well promotes the LV coexistence. In both SWBW cases, the macrophase separation appears for any well depth or barrier height.
In this case it is clearer that the second well presence not only increases an effective interaction range, but also an effective depth well.

%%%%%%%%%%%%%%%%%%%%%%%%
% Figure 8
%%%%%%%%%%%%%%%%%%%%%%%%     

Finally, we present a data table with the critical values found for each fluid.

%%%%%%%%%%%%%%%%%%%%%%%
%    Tabla 2
%%%%%%%%%%%%%%%%%%%%%%5
\begin{center}
  \begin{tabular}{| l | c | r | l | c | r | l | c | r | l | }
    \hline
   Type & $\epsilon_{2}$ & $\epsilon_{3}$ & $\lambda$ & $\lambda_{2}$&$\lambda_{3}$& $T_{c}^{*}$&$\rho_{c}^{*}$ \\ \hline
     SW   &  0.0 & 0.0 & 1.5 & 2.0 & 2.5 & 1.226 & 0.304   \\ \hline
     SW   &  0.0 & 0.0 & 2.0 & 2.0 & 2.5 & 2.624 & 0.274   \\ \hline
     SW   &  0.0 & 0.0 & 2.5 & 2.0 & 2.5 & 5.67 & 0.276  \\ \hline
     SW   &  0.0 & 0.0 & 3.0 & 2.0 & 2.5 & 9.980 & 0.255 \\ \hline     
     SWB  &  0.1 & 0.0 & 1.5 & 2.0 & 2.5 & 1.080 & 0.310 \\ \hline
     SWB  &  0.2 & 0.0 & 1.5 & 2.0 & 2.5 & 0.972 & 0.306 \\ \hline
     SWB  &  0.3 & 0.0 & 1.5 & 2.0 & 2.5 & 0.851 & 0.301 \\ \hline
     SWB  &  0.4 & 0.0 & 1.5 & 2.0 & 2.5 & 0.760 & 0.297 \\ \hline
     SWB  &  0.5 & 0.0 & 1.5 & 2.0 & 2.5 & 0.700 & 0.294 \\ \hline
     SWB  &  0.1 & 0.0 & 1.25 & 2.0 & 2.5 &  0.559 &  0.3035 \\ \hline
     SWBW &  0.1 & 0.1 & 1.5 & 2.0 & 2.5 & 1.440 & 0.313  \\ \hline
     SWBW &  0.2 & 0.1 & 1.5 & 2.0 & 2.5 & 1.344  & 0.312 \\ \hline
     SWBW &  0.5 & 0.1 & 1.5 & 2.0 & 2.5 & 1.045  & 0.309  \\ \hline
     SWBW &  0.5 & 0.2 & 1.5 & 2.0 & 2.5 & 1.339  & 0.316  \\ \hline
     SWBW &  0.5 & 0.3 & 1.5 & 2.0 & 2.5 & 1.640  & 0.328 \\ 
    \hline
  \end{tabular}
\end{center}

\section{Structure}

The cluster formation has been intensely studied, as we mentioned above, using short attractive and large repulsive potentials. In this section we study the structural properties of the  SW, SWB and SWBW systems. The structure factor is computed by means of MC simulation in the NVT ensemble and OZ equation. For each case, we have realized a wide study about the structure, such study is realized in the equilibrium region, near  of the critical point; exploring a large range of  values of the parameters that define the interaction potential, i. e., the strength and range of each contribution. The critical points are shown in the table 1.

We must consider two conditions on the structure factor for the cluster formation \cite {AStradner,  AShukla_c1, AStradner2, Bomont}: (1) a low-q peak value located at $q^{*}_{c}$, namely cluster-cluster peak; this one corresponds to a mean distance among clusters $\sim 2\pi/q_{c}$  and characterize the repulsive interaction between them. (2) a principal high-q peak   value, namely monomer-monomer peak, located at $q^{*}_{m}$, corresponding to the mean distance among monomers within a single cluster.

\begin{figure}
  \centering
  \includegraphics[width=10.0cm,height=8.0cm]{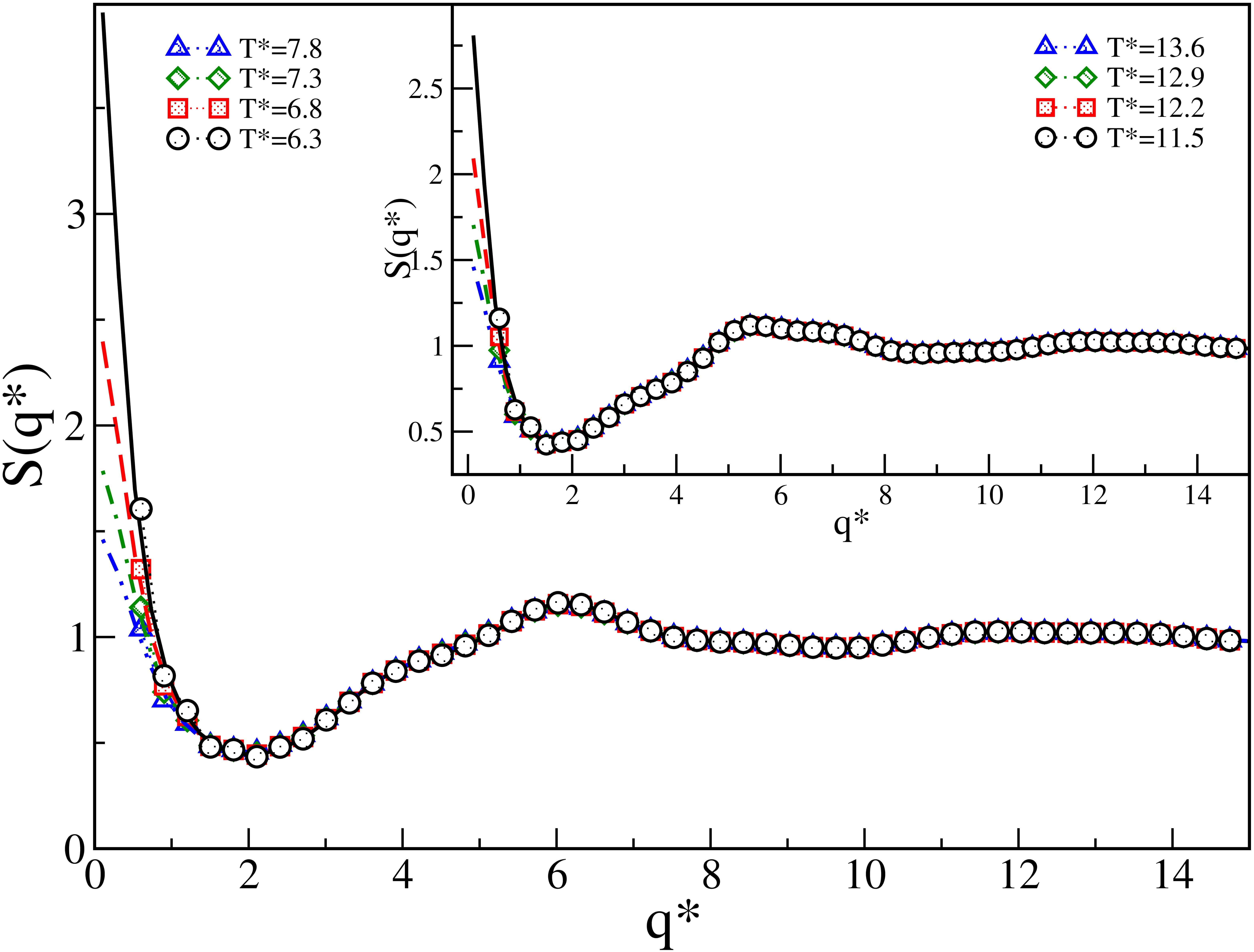}
\caption{{\footnotesize  Structure factor for SW fluids  for $\epsilon=$ 1.0,  $\lambda=$2.5 ,  at $\rho=0.225$. In the inset   $\epsilon=$ 1.0 , $\lambda=$3.0  and $\lambda=$3. Symbols represent  MC simulation data, lines represent OZ-HMSA.}}
\label{figsqsw}
\end{figure}  

For the SW fluid, we analyze the structure for each value of $\lambda$  studied in the last section,  but we show only the cases for $\lambda=$2.5 and 3.0 since these cases have been less studied and  in addition we find an unusual behavior. In figure \ref{figsqsw} we show the structure factor of the SW case for $\lambda=$2.5 and 3.0 (inset). In addition, for  $\lambda=$2.5 and  $\lambda = 3.0$ we observe the formation of a peak near to $q_{m}$ for $q<q_{m}$, for $\lambda=2.5$ the peak is localized at $q_{c} \sim 5.98$ which corresponds to $r \sim 1.05$, and for $\lambda=3.0$ the peak is localized at $q_{c} \sim 5.2$ which corresponds to $r \sim 1.2$. We can observe that for both cases the peak height of the $S(q^{*}_{m})$ is smaller  than the principal peak at $q^{*}_{c}$. Clearly the structure is sensitive to the interaction range and for large values of $\lambda$  a singular behavior appears, we observe that in the system there is a characteristic length $l_{c}=\frac{2 \pi}{q_{c}}$, where $l_{c}> \sigma$. This behavior corresponds to the formation of well-defined domains or dimers, this formation promotes and favors the LV coexistence that is observed to interact at a large range.

\begin{figure}
  \centering
  \includegraphics[width=10.0cm,height=8.0cm]{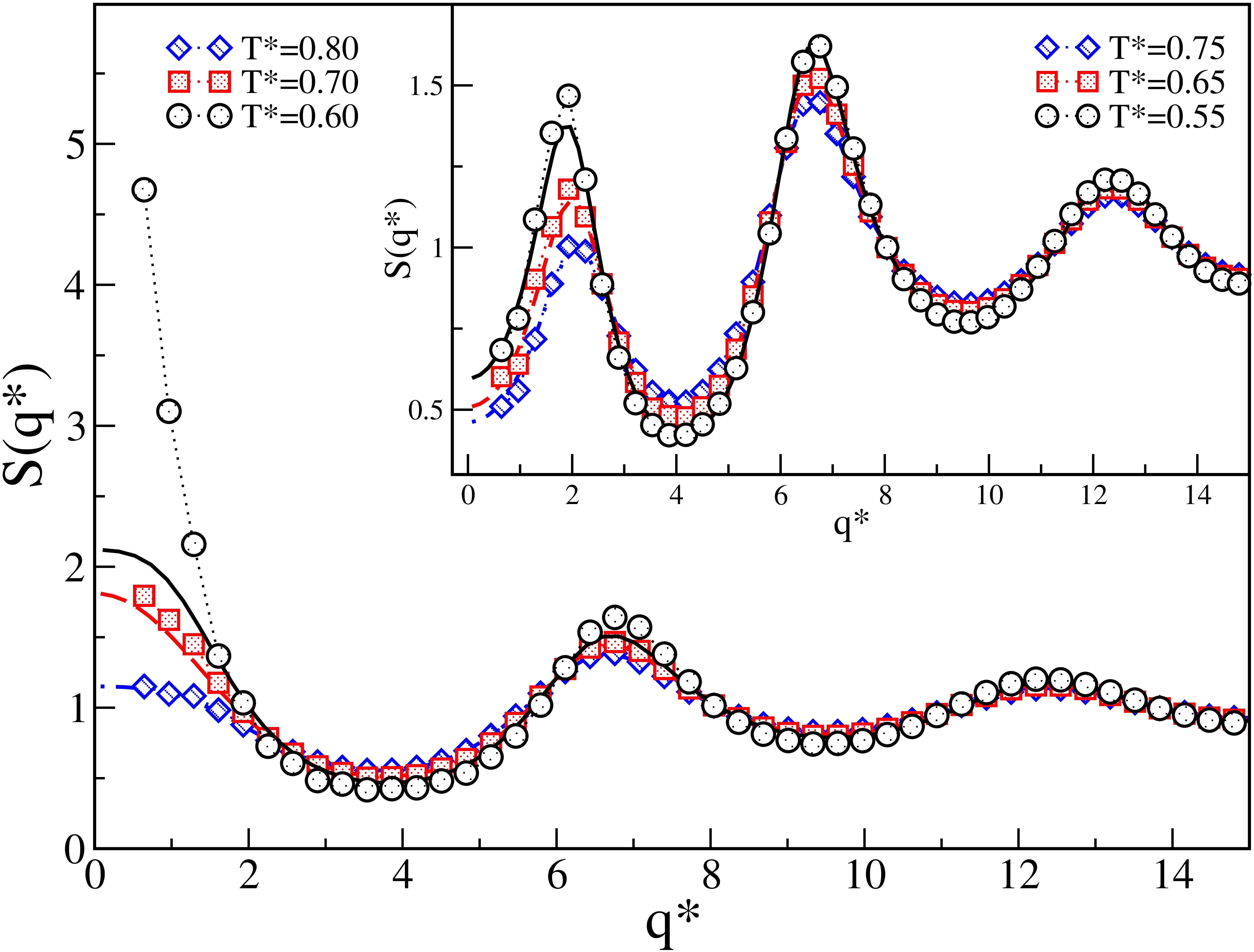}
\caption{{\footnotesize  Structure factor for SWB fluids  for $\epsilon=$ 1.0,  $\epsilon_{2}=$ 0.1, $\lambda=$1.25  and $\lambda_{2}=$2, at $\rho=0.275$.  In the inset $\epsilon=$ 1.0,  $\epsilon_{2}=$ 0.3, $\lambda=$1.25  and $\lambda_{2}=$2. Symbols represent  MC simulation data, lines represent OZ-HMSA and in the outset the solid line is computed by PY.}}
\label{figsqswb1a}
\end{figure}
    
For the SWB fluid, we investigated the same values of $\lambda$ and $\epsilon$ studied in section III. First, we present the results for the short range attractive interaction ($\lambda=1.25$) and a medium range repulsive interaction, where the barrier height is increased. In figure \ref{figsqswb1a} we show the structure  for $\epsilon_{2}=0.1$,  we observe that the value of $S(q\rightarrow 0)$ increases as the temperature decreases, which  is a typical  thermodynamic behavior in liquid-vapor transition. We can say that the thermodynamics mechanism rules the behavior of the system. However, in the inset of figure \ref{figsqswb1a}, where we show the structure factor for $\epsilon_{2}=0.3$, we can observe that at low-q appears a peak and it increases as temperature decreases, this behavior is characteristic of the cluster phase formation. As we saw in the last section, for this system we find LV coexistence for $\epsilon_{2} \le 0.1$, therefore for this interaction ranges, we can affirm that if disappears the LV coexistence due a slight increase of the barrier height, then there are cluster phase formation

\begin{figure}
  \centering
  \includegraphics[width=10.0cm,height=8.0cm]{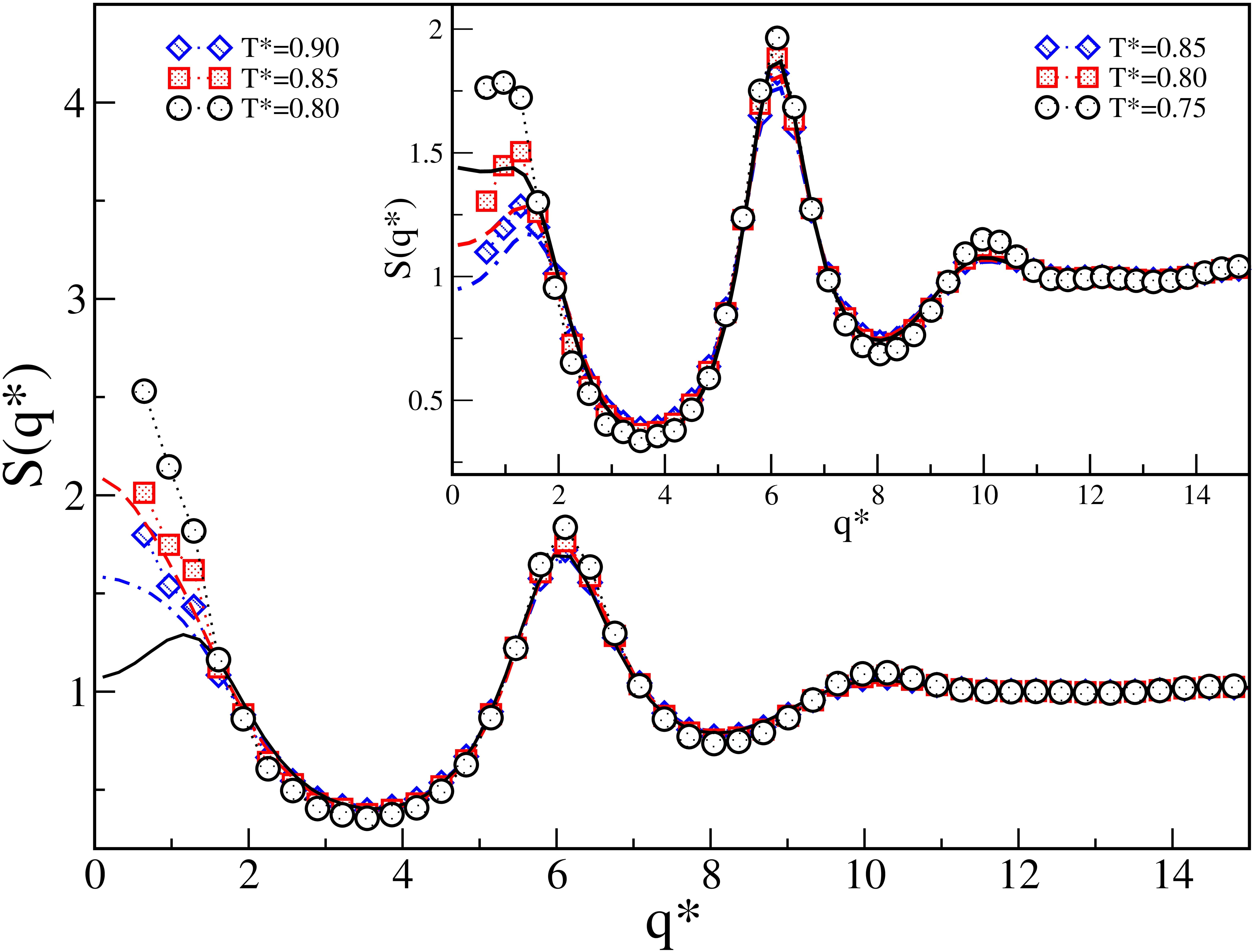}
\caption{{\footnotesize  Structure factor for SWB fluids  for $\epsilon=$ 1.0,  $\epsilon_{2}=$ 0.5, $\lambda=$1.5  and $\lambda_{2}=$2, at $\rho=0.275$.  
In the inset $\epsilon=$ 1.0,  $\epsilon_{2}=$ 0.7, $\lambda=$1.5  and $\lambda_{2}=$2. Symbols represent  MC simulation data, lines represent OZ-HMSA and in both graphics the solid line is computed by PY.}}
\label{figsqswb2a}
\end{figure}

In figure \ref{figsqswb2a}, we show the structure factor for the system with medium-range attractions and  medium-range repulsions  interaction, i.e. $\lambda_{1}=1.5$ and $\lambda_{2}=2.0$, with $\epsilon_{1}=1.0$, $\epsilon_{2}=0.5$ and 0.7 (inset); we explore the cases where $\epsilon_{2} \in [0.1,0.7]$. We find that for $\epsilon_{2}\le 6$ the fluid has a typical behavior, the $S(q\rightarrow 0)$ values increases as the temperature decreases and no low-q peak appear. For $\epsilon_{2}=0.7$ it appears a slight formation of a peak at $q<2$, the peak is more clear as temperature increases. Then  a higher barrier is needed  to find a cluster phase, this coincides with the fact that the LV coexistence disappears for $\epsilon_{2}\approx 0.7$, see the last section.

\begin{figure}[htbp]
  \centering
  \includegraphics[width=10.0cm,height=8.0cm]{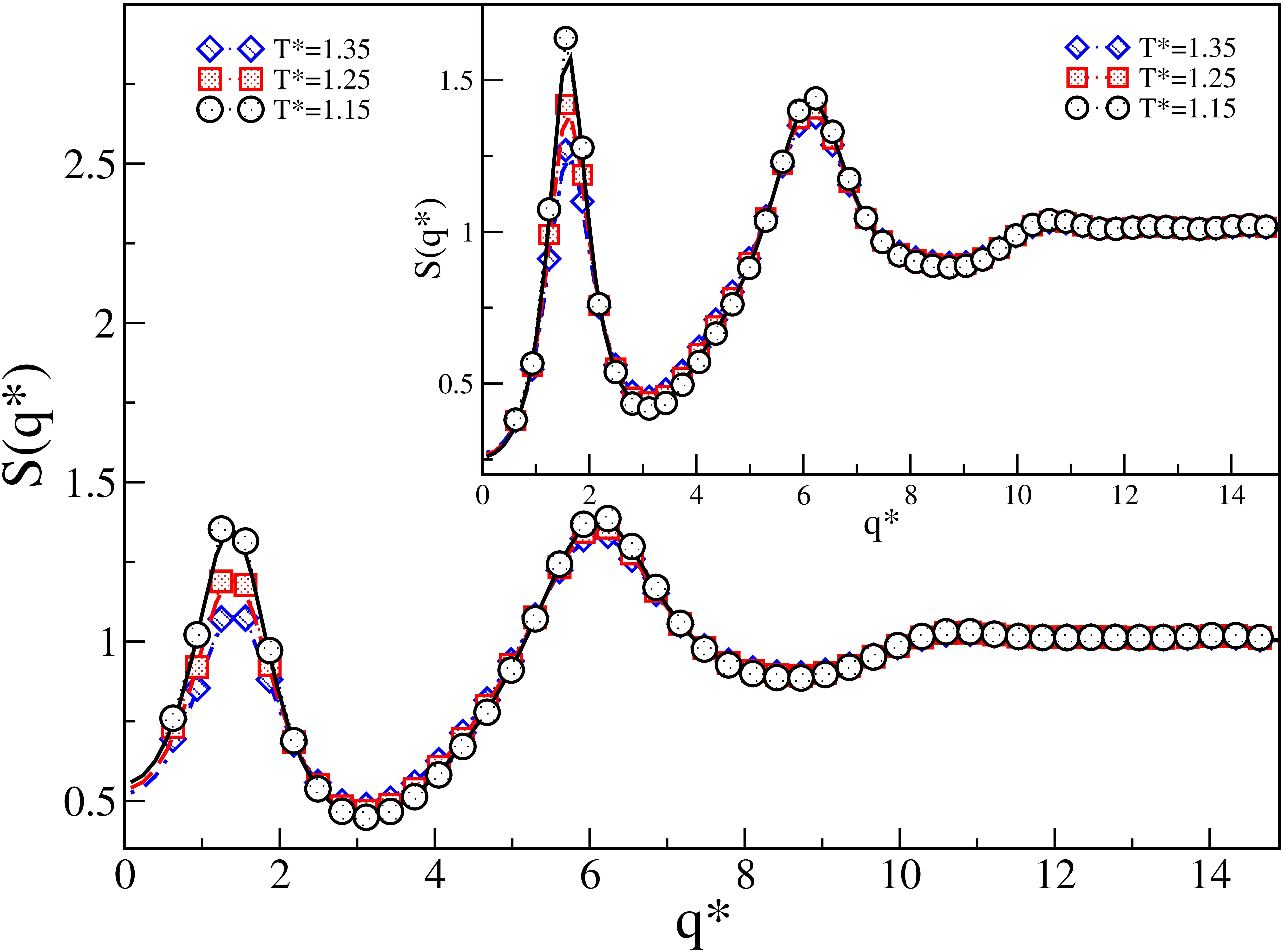}
\caption{{\footnotesize  Structure factor for SWB fluids  for $\epsilon=$ 1.0,  $\epsilon_{2}=$ 0.1 , $\lambda=$1.5  and $\lambda_{2}=$3, at $\rho=0.250$. In the inset   $\epsilon=$ 1.0,  $\epsilon_{2}=$ 0.2 , $\lambda=$1.5  and $\lambda_{2}=$3. Symbols represent  MC simulation data, lines represent OZ-HMSA.}}
\label{figsqswb3a}
\end{figure}   

The last case of SWB fluid is shown in the figure \ref{figsqswb3a}, this case corresponds to medium-range attractions and  long-range repulsions: $\lambda_{1}=1.5$ and $\lambda_{2}=3$, respectively. We explore $\epsilon_{2}$ values in $[0.1,0.5]$ but we show two significative values: $\epsilon_{2}=0.1$ and 0.2 (inset). For this case, we find that the liquid-vapor coexistence disappears even for a small repulsive barrier. We observe the cluster formation for any barrier height  and this is a clear consequence of the large-range repulsive interaction. We can observe a peak for low-q values, i. e. there is a $q_{c}$ that is a characteristic of a cluster formation, this peak appears for any temperature value.

For system SWB exist a clear relation between the cluster phase formation, and both strength and range of the attractions and repulsions.  Within of a comprehensive study, we encounter a qualitative behavior, as it is shown in figure 10; we find the region when the LV transition is inhibited and clusters formation is observed, this correspond to black dots,  the white dots correspond to LV coexistence region of two representative systems: SWB with $\lambda=1.5$, $\lambda_{2}=2.0$ (SWB1) and SWB with  $\lambda=1.25$, $\lambda_{2}=2.0$ (SWB2),  the critical temperatures of these systems are shown in Table1,  in cases where we detected micro-phase (black dots) we explored various temperatures like in the study of structure factors,  that is, near and below the critical temperature of a similar system with LV coexistence (figure 7 y 8). After this study,  we set an exponential fit for the boundary between both regions, above this curve there are clusters formation; We take as example SWB1 and SWB2:  When $\lambda_{r}=0.33$, as in the case SWB1, it need a barrier height greater than $\epsilon_{2}=0.7$ for inhibiting the LV transition; while $\lambda_{r}=0.6$, case SWB2, the LV transition is inhibited with  barrier height $\epsilon_{2}=0.2$ (figure 3), in general, when $\lambda_{r}$ is greater than or equal to 1, there are micro-phase  separation even with barrier height $\epsilon_{2} < 0.1$ (figure 9).

%In general, for the system SWB there are a clear dependence between  the LV coexistence or cluster phase formation, and the barrier height as well as of the interactions ranges; also exist a close relation among two phases. We find a qualitative behavior between this phases, as is showed in the figure \ref{limite}, where the dependence of the cluster formation and of the LV coexistence with respect to the barrier height and interaction ranges difference is sketched.  Within of a comprehensive study, we find a region of LV coexistence as function of $\lambda_{r}$ and $\epsilon_{r}$, in the figure \ref{limite} the withe points correspond to LV coexistence and the black points correspond clusters-phase formation; in the figure is shown only representative datas. After of this study, we set an adjust for the boundary between the regions of cluster formation and of coexistence, such boundary has an exponential form.

\begin{figure}[htbp]
  \centering
  \includegraphics[width=10.0cm,height=8.0cm]{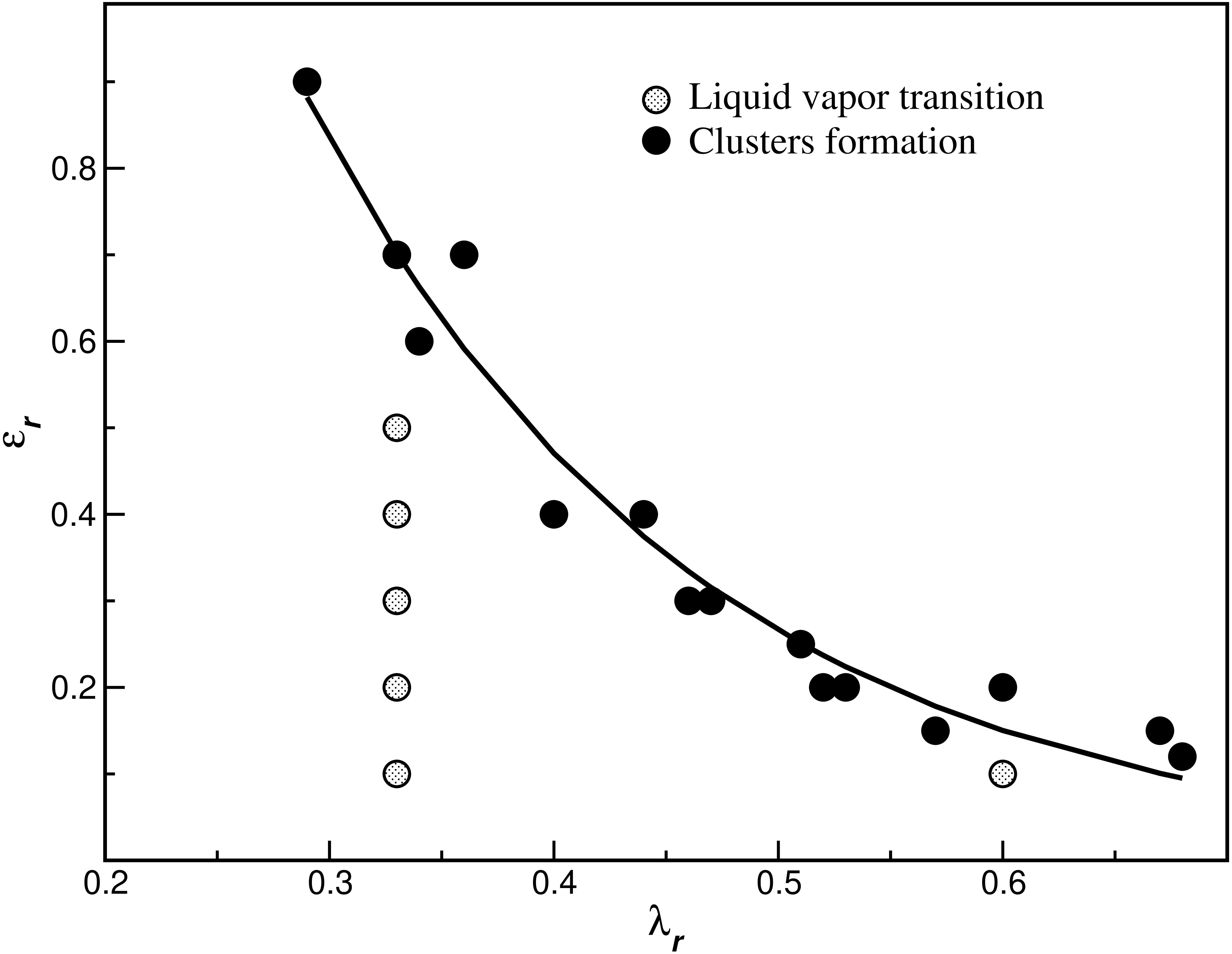}
\caption{{\footnotesize  Schematic diagram of the cluster phase formation and macro-phase separation as function of range and strength of the interaction.}}
\label{limite}
\end{figure} 

\begin{figure}[htbp]
  \centering
  \includegraphics[width=10.0cm,height=8.0cm]{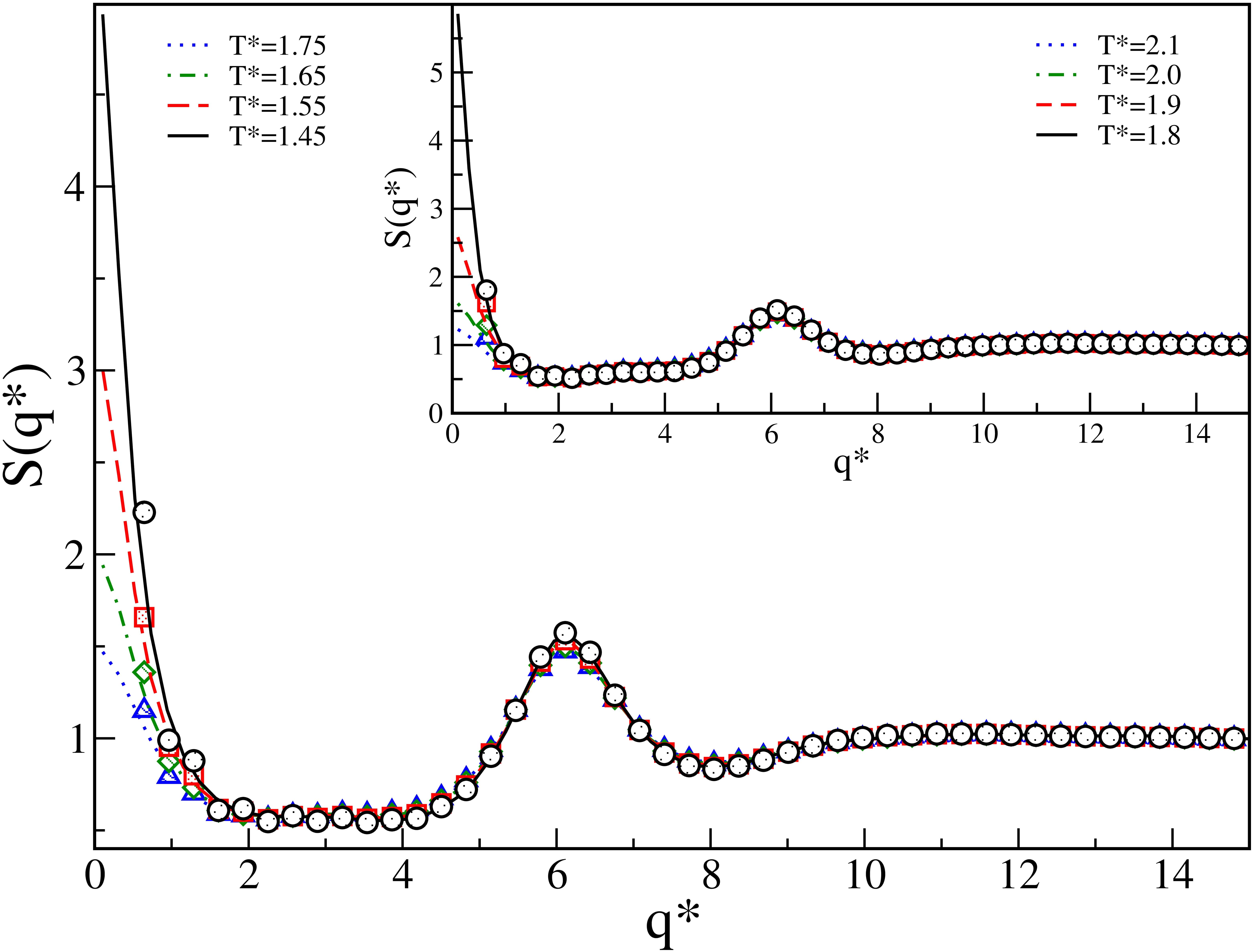}
\caption{{\footnotesize  Structure factor for SWBW fluids  for $\epsilon=$ 1.0,  $\epsilon_{2}=$ 0.5,  $\epsilon_{3}=0.3$,$\lambda=$1.5, $\lambda_{2}=$2 and $\lambda_{3}=2.5$ at $\rho=0.275$. In the inset   $\epsilon=$ 1.0,  $\epsilon_{2}=$ 0.3, $\epsilon_{3}=0.2$ $\lambda=$1.5 , $\lambda_{2}=$2 and $\lambda_{3}=2.5$. Symbols represent  MC simulation data, lines represent OZ-HMSA and in both graphics the solid line is computed by PY.}}
\label{figsqswbw1a}
\end{figure}

Finally, we discuss the SWBW case to complete our study on the structure of a discrete potential fluid,  a second attractive contribution is added to the SWB case, we consider a medium-range interaction: $\lambda=1.5$, $\lambda_{2}=2.0$, $\lambda_{3}=2.5$. We study the competing effect on the structure properties, in the same way that in the study of LV coexistence: the interaction ranges is fixed, and 
the strenght of the repulsive  and second attractive interaction are varied. 

\begin{figure}[htbp]
  \centering
  \includegraphics[width=10.0cm,height=8.0cm]{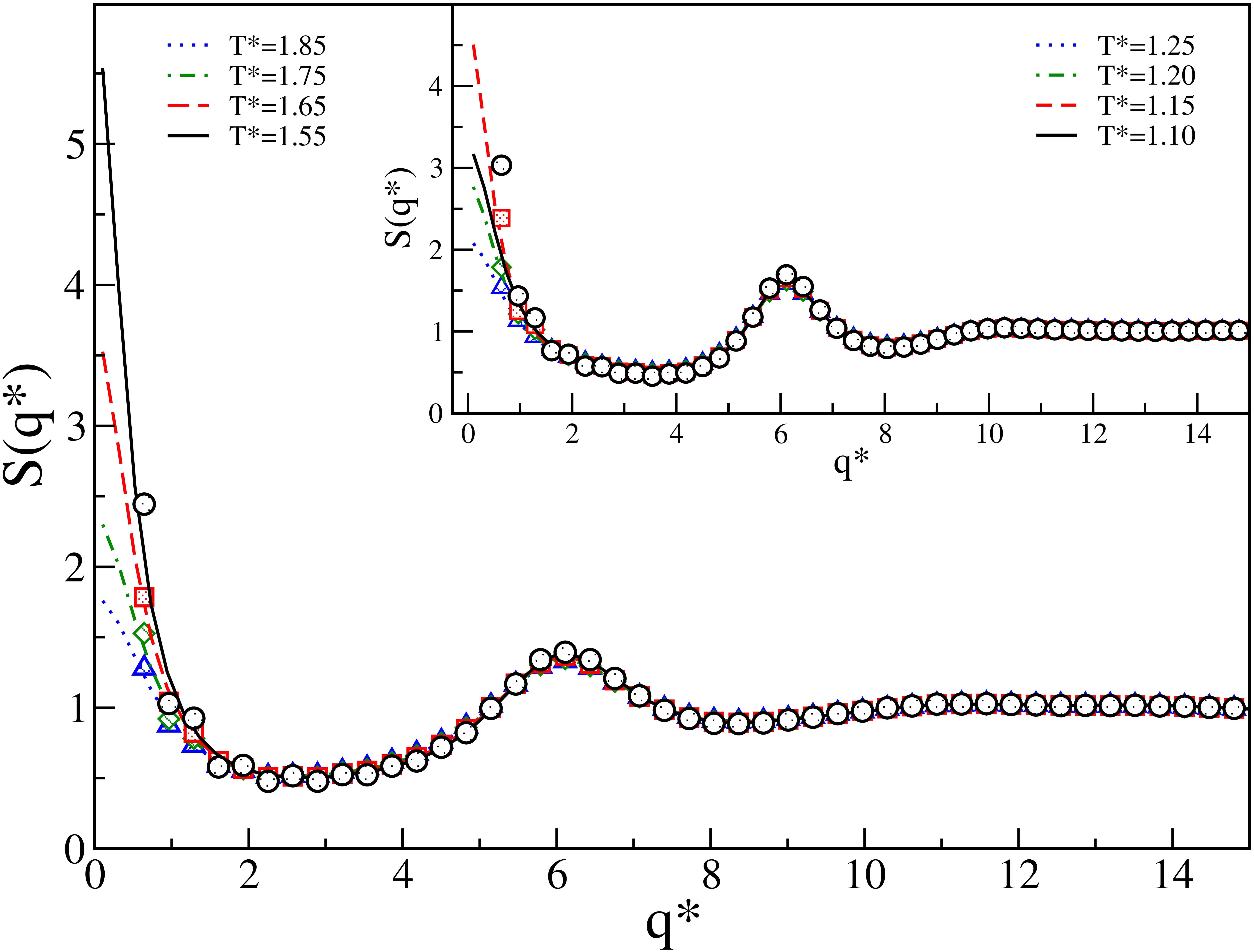}
\caption{{\footnotesize  Structure factor for SWBW fluids  for $\epsilon=$ 1.0,  $\epsilon_{2}=$ 0.3, $\epsilon_{3}=0.1$ $\lambda=$1.5 , and $\lambda_{2}=$2 and $\lambda_{3}=2.5$, at $\rho=0.275$. In the inset   $\epsilon=$ 1.0,  $\epsilon_{2}=$ 0.1, $\epsilon_{3}=0.1$ , $\lambda=$1.5 , $\lambda_{2}=$2 and $\lambda_{3}=2.5$. Symbols represent  MC simulation data, lines represent OZ-HMSA and in the inset  the solid line is computed by PY.}}
\label{figsqswbw2a}
\end{figure}

In the first case, the depth of attractive contribution is fixed and the height of the repulsive contribution is varied. In the second case the second attractive contribution is varied  and the other contributions are fixed. For both cases we find a typical  behavior of fluid phase, in the figures \ref{figsqswbw1a} and \ref{figsqswbw2a} the corresponding structure factors are showed. The presence of the second well promotes the macrophase separation regardless of the depth of the second well, which coincides with the results in the previous section. For this cases,  the effects induces by the barrier for favoring the micro-phase separation vanish.

Within the study of the structure, we use a wide number of closure relations to solve OZ equation: PY, MSA, HNC, HMSA, RY; the last two are thermodynamically self-consistent. 

We find what closure relation is the best for each interaction potential according to the density near the coexistence region. For the SW and SWBW cases, the RY closure is not a good approximation to solve the OZ equation, since it is not convergent for most of the density values, this agrees with a previous work \cite{ALang}.  MSA closure converges in most cases but the results do not agree with the simulation data.  On the other hand, for temperatures above the critical point PY, HNC and HMSA converge and the results agree with MC data. But near of the critical point HNC and HMSA not converge. 

For the fluid SWB we find a similar scene in the case where the coexistence LV appear, i.e., around the critical point only PY converge but it is not a good approximation; and above the critical point PY, HNC and HMSA converge and they become the best approximations, but for $q \rightarrow 0$ differences between the results of the different closures appear, without knowing what  the best in comparison with the simulation is. In the cases where a phase diagram do not arise, the three closure relations converge and at low temperatures there are differences being PY, HMSA and HNC the approximations that reproduce the simulation data, in particular they reproduces the second peak at q-low.

%In the SWBW case we find that as in the SW case, RY and HNC closures is not a good approximation since not converge near of critical point or not reproduce the MC data, PY closure converge near of the critical point but no is accord with  MC data at q-low. However the HMSA closure converges and agrees with the simulation results. 

\section{Conclusions}

With no competing interaction potential, like the SW, the interaction range plays a main role in the micro- and macro-phase separation. On the one hand, it is known that the attractive contribution of the potential promotes the LV coexistence and if the interaction range increases then the coexistence region appears at lager temperatures. On the other hand, we find that the increase of the interaction range favors the micro-phase formation, specifically  for $\lambda \ge 2.5$ we find that a domains or dimers formation appears.

Competing interaction, SWB type, can give rise to micro- and macro-phase behavior that can be different from the one found in simple fluids. In this case, not only the interaction range plays a main role in the micro- and macro-phase separation, but also the interaction strength. We find that for short-range attractive interaction  and long-range repulsive interaction, the LV coexistence is inhibited and a micro-phase separation appears when the barrier height is increased, it is a fact that the cluster formation depends on the interaction range and on the interaction strength. In addition, we observe that the poor LV coexistence is related with the cluster-phase formation. This same effect is shown in the case where the attractive interaction is medium-range and the repulsive interaction is large-range, but in this case the cluster phase formation is independent of the barrier height. In both SWB cases, the cluster formation is related to the absence of the LV coexistence. However for attractive and repulsive interactions of medium-range it is needed a larger barrier height to observe a trace of cluster phase formation, the macro-phase separation dominates in this case, the LV coexistence appears for several values of the barrier height.

Finally, it is clear that the second well favors the macro-phase separation, the LV coexistence appears regardless the barrier height. The attractive interaction of long-range inhibits the cluster phase formation.\\

\end{document}